\documentclass[aps,prd,amsmath,amssymb,preprintnumbers,nofootinbib,a4paper,11pt]{revtex4-1}
\pdfoutput=1
\usepackage{amsthm}
\usepackage{graphicx}
	\graphicspath{{./NewFig/}}
\usepackage{url}
\usepackage[bookmarks, pagebackref=false]{hyperref}
\usepackage{color}
	\definecolor{rossoCP3}{cmyk}{0,.88,.77,.40}
		\definecolor{graa}{rgb}{0.8,0.8,0.8}
		\definecolor{blaa}{rgb}{0.2,0.2,0.6}
		\hypersetup{
			colorlinks,
			bookmarksopen,
			bookmarksnumbered,
			citecolor=blaa, 		
			linkcolor=rossoCP3,	
			urlcolor=rossoCP3,			
			}
\usepackage{dcolumn}
\usepackage{bm}
\usepackage{bbm}
\usepackage{pxfonts}

\usepackage{epsfig}
\usepackage{placeins}

\usepackage[ margin=5pt, font=small,labelfont=bf, justification=raggedright]{caption}
\usepackage{youngtab}
\usepackage{slashed}
\usepackage{subfig}

\newcommand{\beq}{\begin{eqnarray}}
\newcommand{\eeq}{\end{eqnarray}}

\newcommand{\bmp}{\noindent\begin{minipage}{16cm}}
\newcommand{\emp}{\end{minipage}\vskip 7mm} 

\def\lsim{\mathrel{\rlap{\lower4pt\hbox{\hskip1pt$\sim$}}
    \raise1pt\hbox{$<$}}}                
\def\gsim{\mathrel{\rlap{\lower4pt\hbox{\hskip1pt$\sim$}}
    \raise1pt\hbox{$>$}}}                

\begin{document}
\title{\Large  \color{rossoCP3}  Safe Glueballs and Baryons}
\author{Thomas A. Ryttov$^{\color{rossoCP3}{\varheartsuit}}$}\email{ryttov@cp3.sdu.dk}
\author{Kimmo Tuominen$^{\color{rossoCP3}{\clubsuit}}$}\email{kimmo.i.tuominen@helsinki.fi}
 \affiliation{
$^{\color{rossoCP3}{\varheartsuit}}${ \color{rossoCP3}  \rm CP}$^{\color{rossoCP3} \bf 3}${\color{rossoCP3}\rm-Origins},
University of Southern Denmark, Campusvej 55, 5230 Odense M, Denmark
}
\affiliation{$^{\color{rossoCP3}{\clubsuit}}$  \mbox{Department of Physics \& Helsinki Institute of Physics}\\
\mbox{P.O. Box 64, FI-00014 University of Helsinki}}
\begin{abstract}
We consider a non-Abelian gauge theory with $N_f$ fermions and discuss the possible existence of a non-trivial UV fixed point at large $N_f$. Specifically, we study the anomalous dimension of the (rescaled) glueball operator $\text{Tr}\ F^2$ to first order in $1/N_f$ by relating it to the derivative of the beta function at the fixed point. At the fixed point the anomalous dimension violates its unitarity bound and so the (rescaled) glueball operator is either decoupled or the fixed point does not exist. We also study the anomalous dimensions of the two spin-$1/2$ baryon operators to first order in $1/N_f$ for an $SU(3)$ gauge theory with fundamental fermions and find them to be relatively small and well within their associated unitarity bounds.
\end{abstract}

\preprint{HIP-2019-6/TH}

\maketitle

\section{Introduction}

Gauge theory with many flavors of fermions may possess an ultraviolet (UV)
fixed point. This was first studied in Abelian gauge theory
in~\cite{PalanquesMestre:1983zy} and in the non-Abelian case
in~\cite{Gracey:1996he}. Recently, there has been a revived interest in
asymptotically safe four-dimensional quantum field theories, with
particular emphasis on pure fermionic gauge theories, chiral gauge theories, gauge-Yukawa
theories involving both scalars and
fermions and theories with or without supersymmetry \cite{Holdom:2010qs,Shrock:2013cca,Litim:2014uca,Litim:2015iea,Antipin:2018zdg,Molgaard:2016bqf,Antipin:2017ebo,Dondi:2019ivp,Bond:2016dvk,Bond:2017lnq,Bond:2017suy,Bond:2017tbw,Litim:2018pxe,Bond:2018oco,Esbensen:2015cjw,Alanne:2018ene,Alanne:2018csn,Kowalska:2017pkt}. For a recent review of the technical and computational aspects see \cite{Gracey:2018ame}.

The earlier works~\cite{Sannino:2004qp} on fermionic gauge theory phase diagrams as a function of the number of colors, fermion flavors and their representations were
extended to address also the domain where a UV fixed point might emerge in~\cite{Antipin:2017ebo}. While the studies of an infrared (IR) fixed point have been studied both analytically and numerically, the corresponding methodology to study UV fixed points is currently developing. The work reported in this paper is part of this ongoing effort to elucidate the phase structure of gauge theories at large $N_f$ and should be contrasted to other analytical calculations as well as future lattice simulations.

In this paper we will first review the results on the exact gauge theory beta function to leading order in $1/N_f$. Our main results are the formulas for
the anomalous dimensions of the glueball operator and for the two spin-$1/2$ baryon operators to leading order in $1/N_f$. We will then determine the location of the UV fixed point and evaluate the anomalous dimensions at the fixed point.

We find that while the anomalous dimensions of the baryon operators have small values at the fixed point, well consistent with the unitarity bound, the anomalous dimension of the glueball operator grows rapidly with increasing $N_f$ grossly violating its unitarity bound. We will discuss the interpretation of this result. One possible interpretation is that the leading order beta function is inconsistent and there is no fixed point. However, an equally viable interpretation is that the glueball operator is decoupled and not part of the spectrum.

The paper is organized as follows: In section~\ref{sec:intro} we present the
the leading order beta function in detail. Then, in section~\ref{sec:opdims} we
derive the general results for the operator dimensions of the glueball operator and baryon spin-$1/2$ operators, and in section~\ref{sec:uvfp} we analyze
these operator dimensions at a UV fixed point. In section~\ref{sec:checkout} we present our conclusions and outlook for further work.

\section{The beta function at large $N_f$}
\label{sec:intro}

We consider a fermionic gauge theory with gauge group $G$ and $N_f$ number of Dirac fermions in some representation $r$ of $G$.  We let $T_r$ denote the trace normalization factor and $C_r$ the quadratic Casimir of the generators in the representation $r$ while $A$ denotes the adjoint representation. As an expansion in the gauge coupling $\alpha = \frac{g^2}{4\pi}$ we write the beta function as
\begin{eqnarray}
\beta(\alpha) = \mu \frac{d  \alpha}{d  \mu} = - \sum_{i=1}^{\infty} b_i  \frac{\alpha^{i+1}}{\pi^i} = - b_1 \frac{\alpha^2}{\pi} - b_2 \frac{\alpha^3}{\pi^2} - b_3 \frac{\alpha^4}{\pi^3} - \ldots
\end{eqnarray}
Within the class of mass independent schemes the first two coefficients are universal whereas from higher orders they become scheme dependent. Currently the first five coefficients of the beta function in the $\overline{\text{MS}}$ scheme are known \cite{Baikov:2016tgj,Herzog:2017ohr}. These coefficients generally depend on various (higher order) group invariants, the Riemann zeta function $\zeta_s$ and rational numbers.

All coefficients $b_i,\ i\geq 1$ are polynomials in  $N_f$. The first coefficient $b_1$ is a polynomial in $N_f$ to ${\cal{O}}(N_f)$ while all the remaining coefficients $b_i,\ i\geq 2$ are polynomials in $N_f$ to order ${\cal{O}}(N_f^{i-1})$. For these higher order coefficients we will then write
\begin{eqnarray}
b_i &=& b_{i,0} + b_{i,1}N_f + \ldots + b_{i,i-1}N_f^{i-1} = \sum_{j=0}^{i-1} b_{i,j}N_f^{j} \ , \qquad i \geq 2.
\end{eqnarray}

Instead of expanding in the gauge coupling $\alpha$ it is possible to reformulate perturbation theory as an expansion in $1/N_f$ with the normalized coupling $A \equiv T_r  N_f\frac{ \alpha}{\pi}$ held fixed \cite{PalanquesMestre:1983zy,Gracey:1996he,Holdom:2010qs}. Here one rewrites the beta function of the gauge coupling by switching to the normalized coupling $A \equiv T_r  N_f\frac{ \alpha}{\pi}$ and then collecting terms in powers of $1/N_f$. One can then study the theory in terms of perturbation theory in $1/N_f$. Changing variables, we find that the beta function can instead be written as
\begin{eqnarray}\label{eq:beta}
\beta(A) = \mu \frac{d A}{d \mu} = \frac{2}{3} A^2 \left( 1+ \sum_{i=1}^{\infty} \frac{H_i(A)}{N_f^i}  \right).
\end{eqnarray}
The functions $H_i(A)$ depend on the scaled coupling $A$ and are directly related to the coefficients of the original beta function. Explicitly, they are given by
\begin{eqnarray}
H_1(A) &=& -  \frac{11}{4} \frac{C_A}{T_r} - \frac{3}{2} \sum_{j=2}^{\infty} \frac{b_{j,j-1}}{T_r^j} A^{j-1}, \\
H_i(A) &=& - \frac{3}{2} \sum_{j=i}^{\infty} \frac{b_{j,j-i}}{T_r^{j}} A^{j-1} \ , \qquad i \geq 2.
\end{eqnarray}
Remarkably, the first term $H_1(A)$ has been calculated to all orders in $A$ so that the beta function is known exactly to first non-trivial order in $1/N_f$.  It was calculated for QED in \cite{PalanquesMestre:1983zy} while for QCD it was calculated in \cite{Gracey:1996he}. Using the notation of \cite{Holdom:2010qs} it is explicitly given by
\begin{eqnarray}
H_1(A) &=& - \frac{11}{4} \frac{C_A}{T_r} + \int_0^{\frac{A}{3}} I_1(x) I_2(x) dx, \\
I_1(x) &=& \frac{(1+x)(2x-1)^2(2x-3)^2\sin(\pi x)^3 \Gamma(x-1)^2 \Gamma(-2x)}{(x-2)\pi^3}, \label{I1} \\
I_2(x) &=& \frac{C_r}{T_r} + \frac{(20-43x+32x^2 -14x^3+4x^4)}{4(2x-1)(2x-3)(1+x)(1-x)} \frac{C_A}{T_r}. \label{I2}
\end{eqnarray}

\begin{figure}
\includegraphics[width=.49\textwidth]{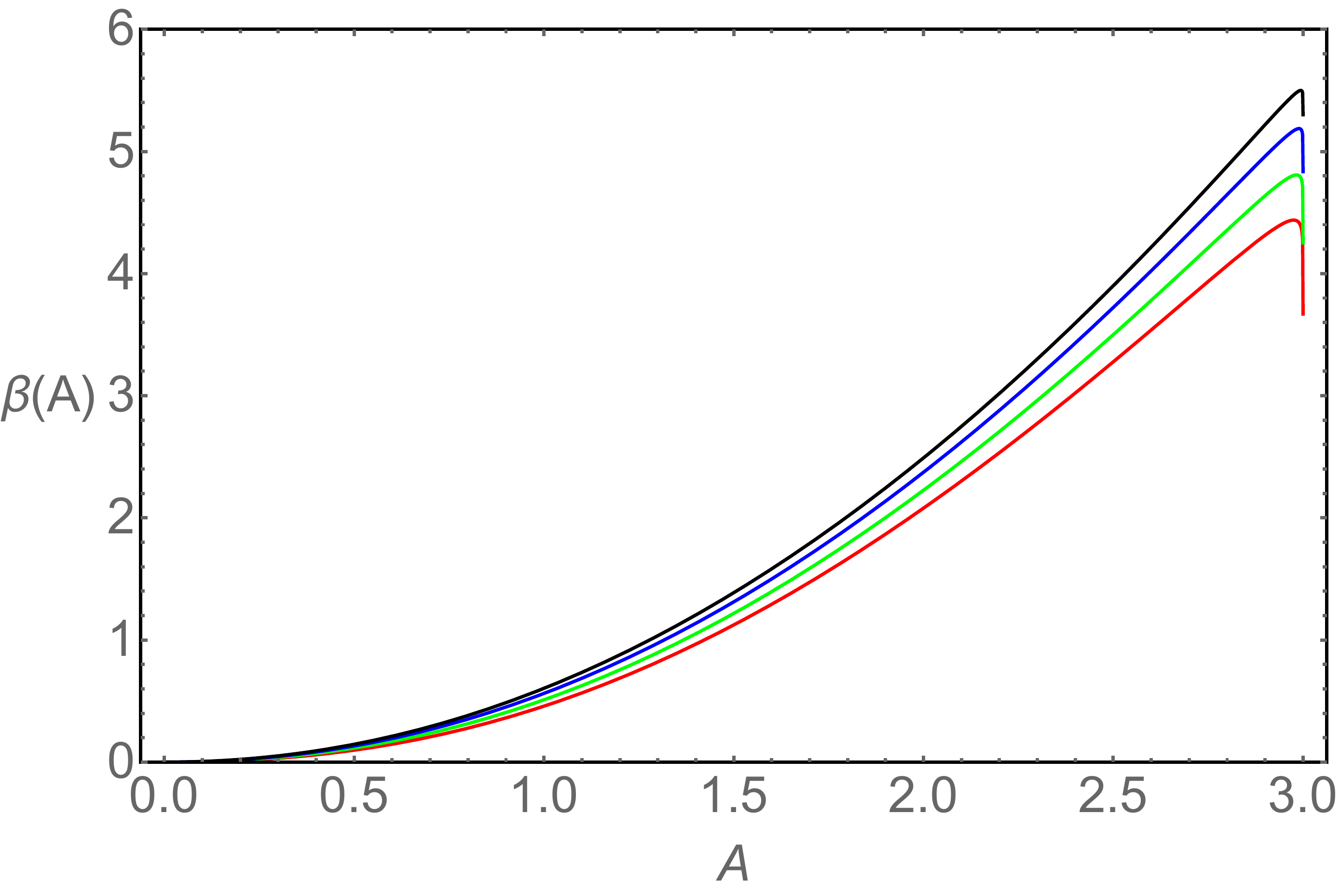}
\includegraphics[width=.49\textwidth]{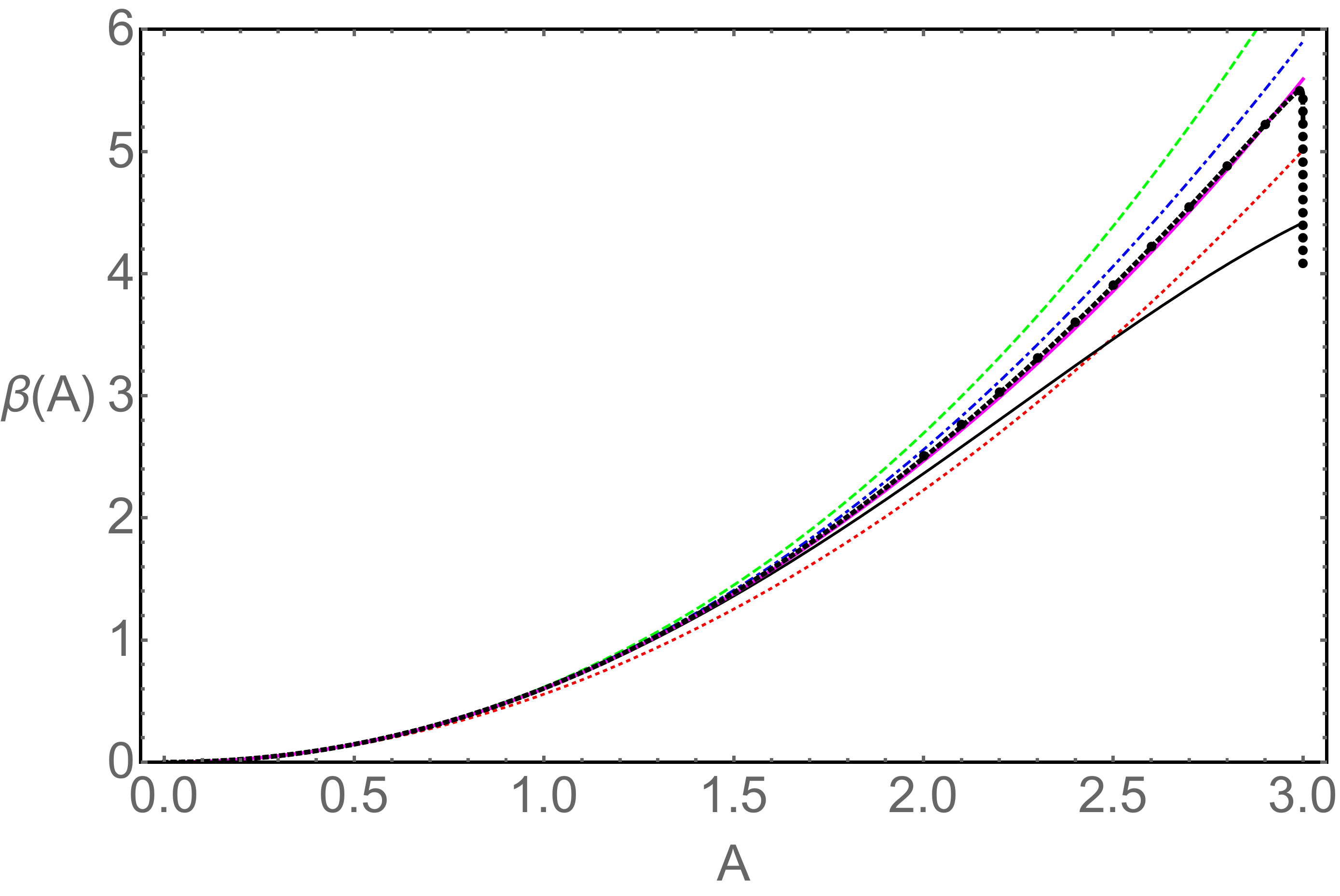}
\caption{Left: The beta function to first order in $1/N_f$ for matter fields in the fundamental representation and $N_c=3$. From bottom to top the curves are for $N_f=30$ (red), $N_f=40$ (green), $N_f=60$ (blue) and $N_f=100$ (black).
Right: The beta function at different loop orders of perturbation theory for $N_f=100$. The dotted (red), dashed (green), dash-dotted (blue), solid (black) and thick solid (magenta) lines correspond, respectively to one, two, three, four and five loop results. The thick-dotted line is the beta function to leading order in $1/N_f$.}
\label{beta_plot}
\end{figure}

In the left panel of Figure \ref{beta_plot} the beta function for fundamental matter and $N_c=3$ to first order in $1/N_f$ is shown for a range of values of $N_f$. In the right panel of Figure~\ref{beta_plot} we show the
comparison of the beta function to leading order in $1/N_f$ and the perturbative
beta function at different loop orders for $N_f=100$. Interestingly, the leading order beta function and five loop beta function agree very well almost up to $A=3$ where the leading order beta function diverges.

At higher orders the functions $H_i(A)$ are not known exactly except for the first few orders in $A$. Specifically the first few terms in $H_2(A),\dots,H_5(A)$ can be inferred from the five-loop beta function and we give explicit expressions for a general gauge group and fermion representation in Appendix~\ref{app:pertH}.
Useful relations between the Euler $\Gamma (z)$, the digamma function $\psi(z)$, the polygamma function $\psi^{(m)} (z)$ and the Riemann zeta function $\zeta(z)$ are provided in Appendix \ref{app:functions}.

\section{Dimensions of Operators at large $N_f$}
\label{sec:opdims}

Of critical importance are the scaling dimensions of gauge invariant operators. At a fixed point they are scheme independent physical quantities. For some gauge invariant operator $O$ we shall in general denote its full scaling dimension by $D = d - \gamma$, where $d$ is its classical scaling dimension and $\gamma$ its anomalous dimension. Note that our convention of the minus sign
if front of the anomalous dimension here. Dimensions of gauge invariant operators have a conventional expansion in the coupling as
\begin{eqnarray}
\gamma &=& \sum_{i=1}^{\infty} \gamma_i \frac{\alpha^i}{\pi^i}.
\end{eqnarray}
Similar to the beta function coefficients the coefficients $\gamma_i$ of the anomalous dimension will in general be a polynomial in the number of flavors $N_f$. Therefore it is also possible to rearrange the perturbative expansion into a $1/N_f$ expansion for the anomalous dimension. In general all of the coefficients of the anomalous dimension can be written as
\begin{eqnarray}
\gamma_i &=& \gamma_{i,0} + \gamma_{i,1} N_f \ldots + \gamma_{i,i-1} N_f^{i-1}  = \sum_{j=0}^{i-1} \gamma_{i,j} N_f^j   \ , \qquad i \geq 1 .
\end{eqnarray}
Note that the first coefficient $\gamma_1$ does not depend on $N_f$ which is different from the first coefficient of the beta function $b_1$. Now again changing variables to the normalized coupling $A = T_rN_f \frac{\alpha}{\pi}$ and collecting terms in powers of $1/N_f$ we can instead write the anomalous dimension as
\begin{eqnarray}
\gamma &=& \sum_{i=1}^{\infty} \frac{G_i(A)}{N_f^i},
\end{eqnarray}
where the functions $G_i(A)$ depend on $A$ and are directly given in terms of the original coefficients of the anomalous dimension as
\begin{eqnarray}
G_i(A) &=& \sum_{j=1}^{\infty} \frac{ \gamma_{j,j-i} }{T_r^j} A^j.
\end{eqnarray}
We will now determine the anomalous dimensions explicitly for the glueball and spin-1/2 baryon operators.

\subsection{Dimension of (rescaled) glueball operator $\text{Tr}\ F^2 $}

If the gauge field is denoted by $A_{\mu}^a$ then the canonical field strength tensor is $F_{\mu\nu}^a = \partial_{\mu} A_{\nu}^a - \partial_{\nu} A_{\mu}^a + g f^{abc} A_{\mu}^b A_{\nu}^c$. If we rescale the canonical gauge fields as $A_{\mu}^a \rightarrow \frac{1}{g} A_{\mu}^a$ then the canonical field strength tensor is rescaled as $F_{\mu\nu}^a \rightarrow \frac{1}{g} F_{\mu\nu}^a$ and the canonical gauge boson kinetic and self interaction term in the Lagrangian is rescaled as $- \frac{1}{4} F_{\mu\nu}^aF^{\mu\nu,a} \rightarrow - \frac{1}{4g^2} F_{\mu\nu}^a F^{\mu\nu,a}$. From now on $F_{\mu\nu}^a$ will denote the field strength tensor rescaled in this way.

The dimension of the $F_{\mu\nu}^a F^{\mu\nu,a}$ operator is of great importance. First, we will show how the dimension of this operator is related to the beta function and its derivative. In order to do this we consider the trace anomaly
\begin{eqnarray}
T^{\mu}_{\phantom{\mu}\mu} &=& \frac{\beta(\alpha)}{16\pi\alpha^2} F_{\mu\nu}^a F^{\mu\nu,a}.
\end{eqnarray}
Taking the derivative on both sides of the trace anomaly by $\mu \frac{d}{d\mu}$ and using the fact that $T^{\mu}_{\phantom{\mu}\mu}$ must scale classically we find that the dimension of the $F_{\mu\nu}^a F^{\mu\nu,a}$ operator must be given by
\begin{eqnarray}
D_{F^2} = 4 + \beta'(\alpha) - \frac{2\beta(\alpha)}{\alpha} \ , \qquad \gamma_{F^2} = - \beta'(\alpha) + \frac{2\beta(\alpha)}{\alpha} = \sum_{i=1}^{\infty} (i-1) b_i \frac{\alpha^i}{\pi^i},
\end{eqnarray}
where $\beta'(\alpha) = \frac{d\beta(\alpha)}{d\alpha}$ is the derivative of the beta function. So the dimension is given directly in terms of the beta function and its first derivative. The coefficients of the anomalous dimension are $\gamma_i  = (i-1)b_i$. Note here that that the first term $\gamma_1 =0$ vanishes and so the first non-trivial contribution to the anomalous dimension begins at order $\alpha^2$.

Using Eq.~\eqref{eq:beta} and switching variable to $A$, we can instead write the dimension as an expansion in $1/N_f$
\begin{eqnarray}
D_{F^2} = 4+ \sum_{i=1}^{\infty} \frac{\frac{2}{3} A^2 H_i'(A)}{N_f^i} \ , \qquad \gamma_{F^2} =  \sum_{i=1}^{\infty} \frac{-\frac{2}{3} A^2 H_i'(A)}{N_f^i}.
\end{eqnarray}
This is one of our main results. It gives the dimension of the rescaled $F_{\mu\nu}^a F^{\mu\nu,a}$ operator to all orders in the coupling and first order in $1/N_f$ since we know $H'_1(A)$ from above. It corresponds to the summation of an infinite set of Feynman diagrams. Explicitly the anomalous dimension to first order in $1/N_f$ is
\begin{eqnarray}
\gamma_{F^2} = -\frac{2}{9} A^2 I_1(A/3) I_2(A/3) \frac{1}{N_f} + {\cal{O}}(1/N_f^2),
\end{eqnarray}
where $I_1$ and $I_2$ are given in Eqs~\eqref{I1}-\eqref{I2}.

\subsection{Dimension of spin-$1/2$ baryon operators $\psi\psi\psi$ in an SU(3) gauge theory}

The second set of operators which we will consider is constituted by the spin-$1/2$ baryon operators $\psi\psi\psi$ for an $SU(3)$ gauge theory with $N_f$ fundamental flavors. Spin-$1/2$ baryon operators come in two types depending on how one contracts the Lorentz indices.  In \cite{Vecchi:2016whd} the anomalous dimensions have been computed exactly to first order in $1/N_f$ for both cases. Due to the presence of evanescent operators the anomalous dimensions contain a function which depends on the scheme in which it is calculated  \cite{Vecchi:2016whd}. We write the full scaling dimension for each of these two operators as
$D_{\pm} = \frac{9}{2} - \gamma_{\pm}$ and using the results of \cite{Vecchi:2016whd} we find
\begin{eqnarray}\label{baryons}
\gamma_{\pm} = \frac{1}{2} \gamma_m \left(\frac{1}{2} A +  \frac{s_{\pm}(A)}{1-\frac{2}{9} A}  \right)   + {\cal{O}}(1/N_f^2).
\end{eqnarray}
Here $s_{\pm}(A)$ are the scheme dependent functions which depend on the coupling $A$ and satisfy $s_{\pm}(0)=1$. We shall consider two different schemes adopted in \cite{Kraenkl:2011qb,Gracey:2012gx} for which $s_{\pm} = 1$ and $s_{\pm} =(1-\frac{2}{9}A )(1-\frac{1}{6}A)$ respectively. Lastly $\gamma_m$ is the anomalous dimension of the mass and has been calculated exactly to second order in $1/N_f$. It was calculated for QCD to first order in \cite{PalanquesMestre:1983zy} and to second order in \cite{Ciuchini:1999wy}. Here we only need it to first order for which it reads
\begin{eqnarray}
\gamma_m &=& \frac{C_r}{2T_r} \frac{A (1-\frac{2}{9} A)  \Gamma(4-\frac{2}{3}A ) }{ \Gamma(1+ \frac{1}{3}A ) [\Gamma (2-\frac{1}{3}A)]^2 \Gamma (3-\frac{1}{3}A) }  \frac{1}{N_f} + {\cal{O}}(1/N_f^2),
\end{eqnarray}
where $T_r = 1/2$ and $C_r = 4/3$ for fermions in the fundamental representation of $SU(3)$. In ordinary perturbation theory it is known to five loop order in the $\overline{\text{MS}}$ scheme \cite{Baikov:2014qja,Baikov:2017ujl}.

\section{Dimensions at the UV fixed point at large $N_f$}
\label{sec:uvfp}

The beta function has a trivial fixed point $(\alpha = 0)$ which is in the IR provided the first coefficient is negative $b_1 <0$, so that the one loop beta function is positive. For a given gauge group the theory is then infrared free if the number of flavors is sufficiently large, $N_f > \frac{11C_A}{4T_r}$. In this region of flavor space it is of considerable interest to ask what happens in the UV. For instance, does the theory run into a Landau pole (which is indicated by the one loop beta function) or does the theory develop a non-trivial UV fixed point. For sufficiently large $N_f$ there are indications that a new non-trivial fixed point is generated \cite{Holdom:2010qs,Shrock:2013cca,Antipin:2017ebo}.

To understand if a non-trivial UV fixed point $A_{UV}$ might be generated we need to study the possible zeros of the beta function. To first order in $1/N_f$ we look for solutions to the equation
\begin{eqnarray}
1+ \frac{H_1(A_{UV})}{N_f} = 0.
\end{eqnarray}
At first sight this might not seem to have any solutions in the large $N_f$ limit. However there are singularities hiding in $H_1(A)$ that can balance $1/N_f$ such as to make the ratio ${H_1(A_{UV})}/{N_f}$ finite and equal to $-1$. Examining the integrand $I_1(x)I_2(x)$ of $H_1(A)$, we see that the first divergence for $x>0$ occurs at $x=1$. This stems from the $\frac{1}{1-x}$ term in $I_2(x)$ and hence gives rise to a logarithmic divergence in $H_1(A)$ at $x=1$. To be more explicit we now expand the integrand in $H_1(A)$ around $x=1$ in order to find
\begin{eqnarray}
I_1(x) I_2(x) = - \frac{C_A}{16T_r} \frac{1}{1-x} + \sum_{i=0}^{\infty} \frac{a_i C_A + b_i C_r}{16T_r} (1-x)^i,
\end{eqnarray}
where $a_i$ and $b_i$ are the expansion coefficients which are pure numbers not depending on any group factors. The first few are $a_0 = 3,\ a_1 =23,\ a_2= 29-\psi^{(2)}(1)$ and  $b_0 = -8,\ b_1= -12, \ b_2 = 32$. This expansion then gives us $H_1(A)$ with the divergence isolated as a logarithm
\begin{eqnarray}
H_1(A) &=& \frac{C_A}{16T_r} \left[ \ln \left( 1- \frac{A}{3} \right)  -a -b \frac{C_r}{C_A} + \sum_{i=0}^{\infty} \left( a_i +  b_i \frac{C_r}{C_A} \right) \frac{\left(1- \frac{A}{3} \right)^{i+1}}{i+1} \right],
\end{eqnarray}
where we have defined $a = 44-\sum_{i=0}^{\infty} \frac{a_i}{i+1}$ and $b=- \sum_{i=0}^{\infty} \frac{b_i}{i+1}$. These two constants cannot be calculated exactly, but using the first 20 expansion coefficients $a_i,\ b_i,\ i=0\ldots,20$ we find that numerically $a= 17.39$ and $b=-5.26$. Both series seem to have safely converged to these values at this order.

We should then look for solutions to $H_1(A_{UV}) = -N_f$ at large $N_f$. Clearly $H_1(A)$ diverges as $A\rightarrow 3$ due to the logarithm term. In this limit, all the terms to the right which are polynomial in $(3-A)$
can be neglected. The resulting equation can be solved for $A_{UV}$, and we find
\begin{eqnarray}
A_{UV} = 3- \delta \ , \qquad \delta = \exp \left[-16 \frac{T_r}{C_A}N_f +18.49 - 5.26\frac{C_r}{C_A} \right].
\end{eqnarray}
This result is in agreement with the results in \cite{Litim:2014uca,Antipin:2017ebo}. We will now assume that this fixed point exists for sufficiently large $N_f$ and study the anomalous dimensions of the glueball operator $\gamma_{F^2}$ and the spin-$1/2$ baryons $\gamma_{\pm}$. At a fixed point the theory is scale invariant and dimensions of gauge invariant operators are scheme independent physical quantities.

The anomalous dimension of the $\text{Tr}\ F^2 $ operator at the UV fixed point in first order in $1/N_f$ becomes
\begin{eqnarray}
\gamma_{F^2} = -\frac{2}{9} A_{UV}^2 I_1(A_{UV}/3) I_2(A_{UV}/3) \frac{1}{N_f} =  \frac{3C_A}{8 T_r} \frac{1}{N_f} \delta^{-1} + {\cal{O}}(\delta^0).
\end{eqnarray}
So the anomalous dimension increases exponentially in the large $N_f$ limit due to the $\delta^{-1}$ contribution. The anomalous dimension is positive (since the derivative of the beta function is negative) and therefore the full dimension $D_{F^2} = 4-\gamma_{F^2}$ blows to minus infinity in the large $N_c$ limit.

If the theory is at a fixed point then there exists unitarity bounds which the scaling dimensions of gauge invariant operators should satisfy. If they do not satisfy this condition then either 1) the theory is not at a fixed point or 2) the operator is not part of the spectrum and has decoupled. Since $\text{Tr}\ F^2$ is a spinless operator the unitarity bound is $D_{F^2} > 1$ \cite{Mack:1975je} which implies that $\gamma_{F^2} < 3 $. This condition is severely violated by many orders of magnitude in the entire range of $N_f$ where the theory is non-asymptotically free $N_f > \frac{11C_A}{4T_r} $.
To see this, we note that $\delta\sim 1$ already for $N_f\sim {\cal{O}}(C_A/T_r)$.

For explicit numbers, consider for example fundamental fermions, $N_c=3$ and $N_f= 17$. Then we find $\gamma_{F^2} \sim 10^{11}$ and increasing exponentially at larger $N_f$. For fundamental fermions, $N_c=2$ and $N_f=12$ we find similarly $\gamma_{F^2} \sim 10^{12}$. For higher representations this becomes even more dire: for adjoint fermions, $N_f=3$ and any $N_c$ we find $\gamma_{F^2} \sim 10^{13}$ and increasing exponentially at larger $N_f$.

The perturbative expansion we are using is of course expected to work only in a subrange of this interval. For fundamental fermions this range was estimated in \cite{Holdom:2010qs} to be for $N_f \gtrsim 10N_c$ while for adjoint fermions this range was estimated in \cite{Antipin:2017ebo} to be for $N_f \gtrsim 7$ for any $N_c$. These are the approximate ranges where the UV fixed point is conjectured to exist and here the value of $\gamma_{F^2}$ is exponentially large. With these considerations in mind we therefore conclude that if the UV fixed point exists then the glueball operator is not part of the spectrum of operators. This should make for interesting lattice simulations to hopefully appear in the near future.

We note that an alternative interpretation of the violation of the unitarity of the anomalous dimension we have described above is that the theory is not at a fixed point. However, in that case one might expect that other observables
would display similar inconsistency. We will now turn to the anomalous dimension of the spin-$1/2$ baryon operators, and show that they remain compatible with the existence of a fixed point.

\begin{figure}
  \begin{center}
    \includegraphics[height=6cm]{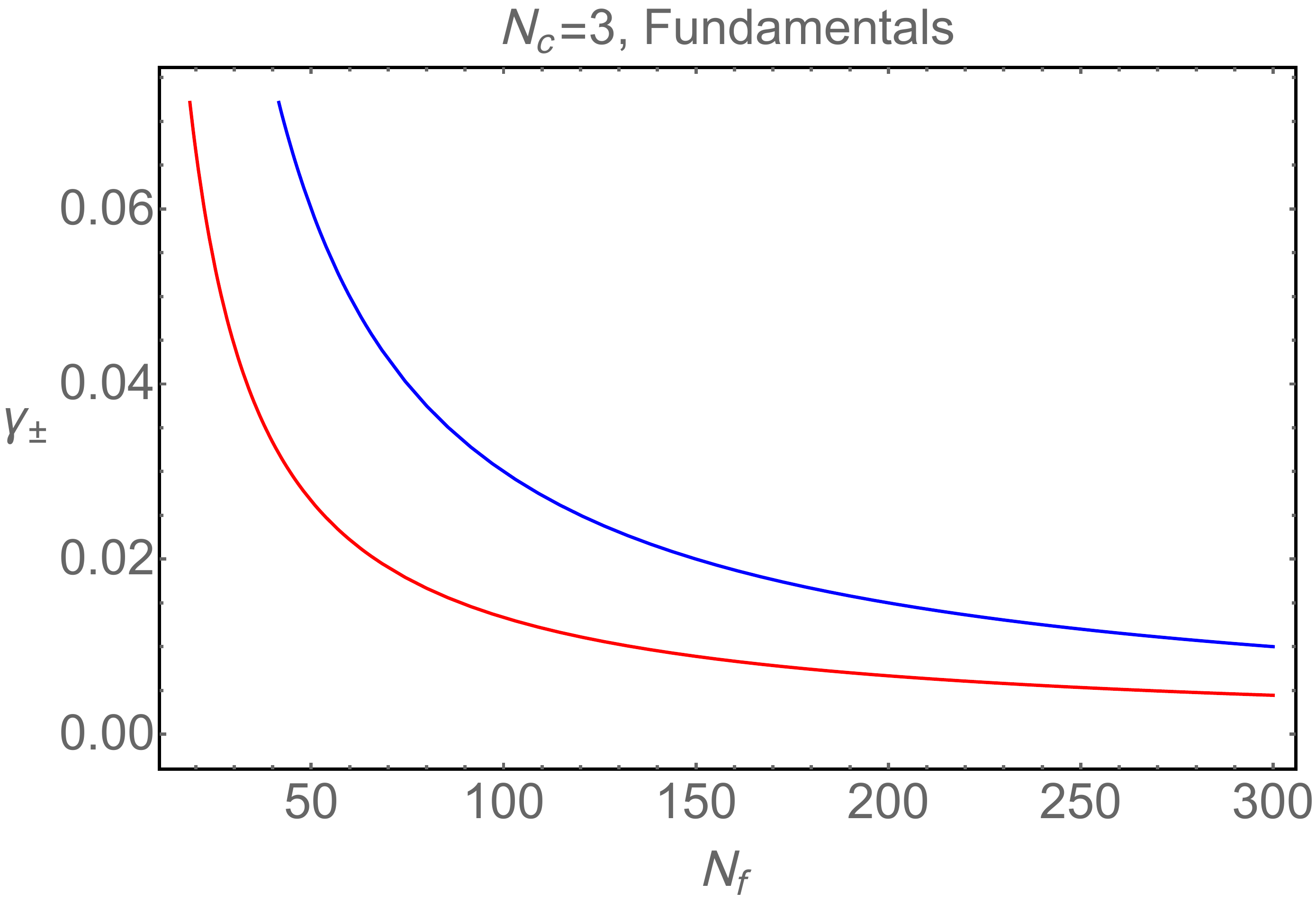}
  \end{center}
\caption{The anomalous dimension of the spin-$1/2$ baryons at the UV fixed point as a function of the number of flavors. The upper blue curve is for the choice $s_{\pm}= 1$ while the lower red curve is for the choice $s_{\pm} = (1-\frac{2}{9}A)(1-\frac{1}{6}A)$.}
\label{baryons_plot}
\end{figure}

Remember, that we are considering two different possibilities for the scheme dependent functions $s_{\pm}$. Evaluating the anomalous dimension at $A_{UV} = 3-\delta$ for the first possibility $s_{\pm}= 1$ we find
\begin{eqnarray}
\gamma_{\pm} = \frac{C_r(3-\delta) (27 +3\delta -2\delta^2) \Gamma\left(2+\frac{2\delta}{3} \right)}{ 72T_r \Gamma \left(2 - \frac{\delta}{3} \right) \Gamma \left( 1+ \frac{\delta}{3}\right)^2 \Gamma \left( 2+ \frac{\delta}{3} \right)} \frac{1}{N_f} \sim \frac{9 C_r}{8T_r}\frac{1}{N_f} + \frac{C_r }{2T_r}\frac{1}{N_f} \delta + {\cal{O}}(\delta^2),
\end{eqnarray}
while for the second possibility $s_{\pm} = (1-\frac{2}{9}A)(1-\frac{1}{6}A)$ we find
\begin{eqnarray}
\gamma_{\pm} = \frac{C_r (3-\delta)(6-\delta) (3+2\delta) \Gamma\left( 2+ \frac{2\delta}{3} \right) }{108T_r \Gamma \left( 2- \frac{\delta}{3} \right) \Gamma \left( 1+ \frac{\delta}{3} \right)^2 \Gamma \left( 2+ \frac{\delta}{3} \right) } \frac{1}{N_f} \sim \frac{C_r}{2T_r} \frac{1}{N_f} + \frac{5C_r}{12T_r} \frac{1}{N_f} \delta + {\cal{O}}(\delta^2).
\end{eqnarray}
For completeness, in Figure \ref{baryons_plot} we show the anomalous dimension as a function of the number of flavors in the entire range $16.5<N_f<300$ where the theory is non-asymptotically free. We remind the reader that the theory is conjectured to develop a non-trivial UV fixed point in the region only in the range $N_f \gtrsim 30$. The upper blue solid curve is for $s_{\pm}= 1$ while the lower red solid curve is for $s_{\pm} = (1-\frac{2}{9}A)(1-\frac{1}{6}A)$.

The value of $\gamma_{\pm}$ is mostly sensitive to the scheme dependent function $s_{\pm}$ in the lower reaches of $N_f$. At $N_f= 30$ we find $\gamma_{\pm} \sim 0.1$ for $s_{\pm} = 1$ and $\gamma_{\pm} \sim 0.04 $ for $s_{\pm} = (1-\frac{2}{9}A)(1-\frac{1}{6}A)$ wheres at $N_f=100$ we find $\gamma_{\pm} \sim 0.03$ for $s_{\pm} = 1$ and $\gamma_{\pm} \sim  0.013$ for $s_{\pm} = (1-\frac{2}{9}A)(1-\frac{1}{6}A)$. However although there is this sensitivity we observe that $\gamma_{\pm}$ is quite small for throughout the entire range $N_f \geq 30$. So with the full scaling dimension being $D_{\pm} = \frac{9}{2} - \gamma_{\pm}$ the spin-$1/2$ baryond operators appear to be irrelevant ($D_{\pm} >4$) for any $N_f$ where the theory is at the UV fixed point never becoming marginal ($D_{\pm} = 4$) or relevant ($D_{\pm} < 4$). The unitarity bound for a spin-$1/2$ baryon operator is $D_{\pm} > \frac{3}{2}$ \cite{Mack:1975je} which is of course also safely satisfied within the entire range $N_f>30$.

\section{Conclusion}
\label{sec:checkout}

In this paper we have considered the possibility of the emergence of
UV fixed points in gauge theory at large number $N_f$ of fermionic matter fields. The beta function is known to leading order in $1/N_f$ and here
we have extended these results to determine the anomalous dimensions of
the glueball and spin-1/2 baryon operators to this same order.

We evaluated the values of these anomalous dimensions at the UV fixed point. For the anomalous dimension of the baryon operators we found that their values are small and well within the unitarity bounds. On the other hand, for the
glueball operator we found that the value of its anomalous dimension increases
rapidly as $N_f$ increases leading to violation of the unitarity bound.
While this result can be interpreted as an inconsistency undermining the existence of the UV fixed point and the validity of the large $N_f$ beta function,
we pointed out that this effect can also arise from decoupling of the glueball
operator from the spectrum at large $N_f$.

The research in this field is currently under active development, and
our results should be useful in further analytic work as well as for
possible lattice simulations of SU($N_c$) gauge theories with many fermion
flavors.

\acknowledgments
TAR wishes to thank N. A. Dondi for useful discussions and the University of Toronto for kind hospitality where the last part of the work was finished. This work is partially supported by the Danish National Research Foundation under the grant DNRF:90 and by the Academy of Finland project 310130.

\appendix

\section{Beta function coefficients}\label{app:pertH}

In this appendix we give the perturbative formulas for the functions $H_i(A)$ appearing in Eq.~\ref{eq:beta} inferred from the perturbative result for
the beta function of generic gauge theory with gauge group $G$ and $N_f$ matter fields in representation $r$ of $G$.
Validating these results requires the use of relations between the Euler $\Gamma (z)$, the digamma function $\psi(z)$, the polygamma function $\psi^{(m)} (z)$ and the Riemann zeta function $\zeta(z)$ which are provided in Appendix \ref{app:functions}.

\begin{eqnarray}
H_1(A) &=& - \frac{11C_A}{4T_r} + \frac{5 C_A+3C_r}{4T_r} A  - \frac{79C_A + 66 C_r}{288 T_r} A^2 - \frac{53C_A + 154 C_r}{2592 T_r} A^3 \nonumber \\
&& + \frac{(-229 + 480\zeta_3 ) C_A + (214 +288\zeta_3) C_r }{20736T_r} A^4  + \ldots
\\
H_2(A) &=& -  \frac{17C_A^2}{8T_r^2} A +  \frac{1415C_A^2 +615 C_A C_r -54 C_r^2}{576T_r^2} A^2 \nonumber \\
&& - \frac{1}{10368 T_r^4} \Bigg( 3(3965+1008\zeta_3)T_r^2C_A^2 + 32(268+189\zeta_3)T_r^2 C_AC_r + 36(169-264\zeta_3)T_r^2 C_r^2  \nonumber \\
&&  +864 (-11 +24 \zeta_3) \frac{d_r^{abcd} d_r^{abcd}}{N_A} \Bigg) A^3   \nonumber \\
&& +\frac{1}{27648 T_r^4} \Bigg( (6231+9736\zeta_3 -3024\zeta_4 - 2880\zeta_5)T_r^2 C_A^2+ 16 (46+1065\zeta_3 -378\zeta_4)T_r^2 C_AC_r    \nonumber  \\
&&   + 2(4961-11424\zeta_3 +4752\zeta_4)T_r^2 C_r^2  + 576 (-55 +123 \zeta_3 -36 \zeta_4 -60\zeta_5) \frac{d_r^{abcd} d_r^{abcd}}{N_A} \Bigg) A^4 +\ldots \nonumber  \\
\\
H_3(A) &=& - \frac{2857C_A^3}{1152T_r^3}A^2 + \frac{1}{20736 T_r^4} \Bigg( -3(-39143 - 3672\zeta_3)T_rC_A^3 + (-7073+17712\zeta_3)T_r C_A^2C_r  \nonumber \\
&&  +36(1051-264\zeta_3) T_r C_AC_r^2 -11178T_rC_r^3 + 3456 (-4 + 39\zeta_3) \frac{d_r^{abcd} d_A^{abcd} }{N_A} \Bigg) A^3 \nonumber \\
&& \frac{1}{165888T_r^5} \Bigg( (-843067-332028\zeta_3 +16848\zeta_4 + 356400\zeta_5) T_r^2 C_A^3  \nonumber \\
&& -3( 5701+158712\zeta_3 - 50976 \zeta_4 +86400\zeta_5 ) T_r^2 C_A^2C_r \nonumber \\
&& -9(94749-57256\zeta_3 +20592 \zeta_4 -79200 \zeta_5)T_r^2 C_AC_r^2 +108 (2509+3216 \zeta_3 -6960\zeta_5) T_r^2 C_r^3 \nonumber \\
&& -864 (-1347+2521\zeta_3 -396 \zeta_4 +140 \zeta_5) C_A \frac{d_r^{abcd} d_r^{abcd}}{N_A} \nonumber \\
&& -51840 (13 +16\zeta_3 -40\zeta_5) C_r \frac{d_r^{abcd} d_r^{abcd}}{N_A} + 1728 (115 -1255\zeta_3 +234\zeta_4 +40\zeta_5) T_r \frac{d_r^{abcd} d_A^{abcd}}{N_A} \Bigg) A^4 \nonumber \\
&& + \ldots
\end{eqnarray}
\begin{eqnarray}
H_4(A) &=& \frac{(-150653 +2376\zeta_3)C_A^4 - 864(-5+132\zeta_3) \frac{d_r^{abcd}d_r^{abcd}}{N_A} }{41472 T_r^4} A^3 \nonumber \\
&& + \frac{1}{20736T_r^5} \Bigg(  3 (39143 - 3672\zeta_3)T_r C_A^3 + ( -7073 + 17712 \zeta_3)T_r C_A^2 C_r + 36 (1051-264\zeta_3)T_r C_A C_r^2 \nonumber \\
&& -11178T_r C_r^3 + 3456 (-4 + 39\zeta_3) \frac{d_r^{abcd} d_A^{abcd} }{ N_A } \Bigg) A^4 + \ldots \\
H_5(A) &=& \frac{(-8296235+78408\zeta_4 + 451440\zeta_5) C_A^5 +2592 (257-9358\zeta_3 +1452\zeta_4 +7700\zeta_5)C_A \frac{d_A^{abcd}d_A^{abcd}}{N_A}  }{1327104T_r^5} A^4 + \ldots \nonumber \\
\end{eqnarray}

\section{The functions $\Gamma (z)$, $\psi(z)$, $\psi^{(m)}(z)$ and $\zeta(z)$}\label{app:functions}

The Euler gamma function $\Gamma(z)$, digamma function $\psi(z)$, polygamma function $\psi^{(m)}(z)$ and Riemann zeta function $\zeta(z)$ are defined as
\begin{eqnarray}
\Gamma(z) &=& \int_0^{\infty} x^{z-1} e^{-x} dx, \\
\psi(z) &=& \frac{d}{dz} \ln \Gamma(z) = \frac{\Gamma'(z)}{\Gamma(z)}, \\
\psi^{(m)}(z) &=& \frac{d^m}{dz^m}  \psi(z) = \frac{d^{m+1}}{dz^{m+1}}\ln  \Gamma (z) \ , \qquad \psi^{(0)} (z) = \psi(z), \\
\zeta(z) &=& \frac{1}{\Gamma(z)} \int_0^{\infty}  \frac{x^{z-1}}{e^x -1} dx-
\end{eqnarray}
A first few examples which might be useful are
\begin{eqnarray}
\Gamma(z+1) &=& z \Gamma(z) \ , \qquad  \psi(z+1) = \psi(z) + \frac{1}{z} \ , \qquad \psi^{(m)}(z+1) = \psi^{(m)} (z) +\frac{(-1)^m m!}{z^{m+1}}  \\
\Gamma(n) &=& (n-1)! \ , \qquad \psi(n) = H_{n-1} - \gamma \ , \qquad \frac{\psi^{(m)}(n)}{(-1)^{m+1} m!} = \zeta(m+1)  - \sum_{k=1}^{n-1} \frac{1}{k^{m+1}}
\end{eqnarray}
where
\begin{eqnarray}
 H_n \equiv \sum_{k=1}^{n} \frac{1}{k}   \ , \qquad \gamma \equiv - \Gamma'(1) = - \psi(1)
\end{eqnarray}
are the Harmonic numbers and the Euler-Mascheroni constant respectively.

\end{document}